\documentclass[11pt,paper]{article}
\usepackage{jheppub}

\usepackage[usenames,dvipsnames,table]{xcolor}
\usepackage{graphicx,amsmath,amssymb,multirow,array,bm,mathrsfs}
\usepackage{epsf,amsfonts}
\usepackage[numbers,sort&compress]{natbib}
\usepackage{dsfont}

\usepackage{slashed}

\newcommand{\be}{\begin{equation}}
\newcommand{\ee}{\end{equation}}
\newcommand{\bea}{\begin{eqnarray}}
\newcommand{\eea}{\end{eqnarray}}

\title{Generalized Gibbs Ensemble and the Statistics of KdV Charges in 2D CFT}

\preprint{TCDMATH 18-17}

\author[a,\circ]{Alexander Maloney}
\note[$\circ$]{maloney@physics.mcgill.ca}
\author[b,c,\diamond]{Gim Seng Ng}
\note[$\diamond$]{gng@math.tcd.ie}
\author[d,\star]{Simon F. Ross}
\note[$\star$]{s.f.ross@durham.ac.uk}
\author[a,\triangleright]{Ioannis Tsiares}
\note[$\triangleright$]{ioannis.tsiares@mail.mcgill.ca}

\affiliation[a]{Department of Physics, McGill University, Montr\'{e}al, QC, Canada}
\affiliation[b]{School of Mathematics, Trinity College Dublin, Dublin 2, Dublin, Ireland}
\affiliation[c]{Hamilton Mathematical Institute, Trinity College Dublin, Dublin 2, Ireland}
\affiliation[d]{Centre for Particle Theory, Department of Mathematical Sciences, Durham University, South Road, Durham DH1 3LE, UK}
\abstract{
Two-dimensional CFTs have an infinite set of commuting conserved charges, known as the quantum KdV charges.  We study the Generalized Gibbs Ensemble with chemical potentials for these charges at high temperature. In a large central charge limit, the partition function can be computed in a saddle-point approximation. We compare the ensemble values of the KdV charges to the values in a microstate, and find that they match irrespective of the values of the chemical potentials. We study the partition function at finite central charge perturbatively in the chemical potentials, and find that this degeneracy is broken. We also study the statistics of the KdV charges at high level within a Virasoro representation, and find that they are sharply peaked. 
}

\begin{document}
\maketitle
\section{Introduction} 
\label{sec:intro}

Two-dimensional conformal field theories (CFTs) have an infinite-dimensional symmetry algebra, the Virasoro algebra, which strongly constrains the structure of the theory. We are interested in exploring the consequences of this symmetry structure for the relationship between individual microstates and thermodynamic ensembles.  A similar investigation has been carried out in \cite{Tolya}, which has some overlap with our analysis. 

In the universal enveloping algebra of the Virasoro algebra one can define an infinite set of commuting charges, which we denote $I_{2m-1}$ \cite{Sasaki:1987mm,Eguchi:1989hs,Bazhanov:1994ft}.  Here $m$ is an integer, and the index $2m-1$ labels the spin of the charge; $I_1$ is the usual Virasoro generator $L_0$. These charges allow us to apply technology from the theory of integrable systems to the Virasoro symmetry structure of two-dimensional CFTs.  The problem of the simultaneous diagonalization of $I_{2m-1}$'s can be mapped to a quantum version of the KdV problem \cite{Bazhanov:1994ft}. Hence the $I_{2m-1}$'s are sometimes called the quantum KdV charges. 

The existence of these charges allows us to introduce a Generalized Gibbs Ensemble (GGE) for two-dimensional CFTs, where we introduce chemical potentials for the KdV charges, 
\begin{equation}
Z[\beta,\mu_3,\mu_5, \ldots] = \mbox{ Tr }[e^{-\beta E + \mu_3 I_3 + \mu_5 I_5 + \ldots}] = \mbox{ Tr }[  e^{ \mu_3 I_3 + \mu_5 I_5 + \ldots} q^{L_0 -k}], \quad q \equiv e^{-\beta}\,,
\end{equation}
where we consider just one chiral sector (there will be an identical structure in the other sector) and use the notation $k = \frac{c}{24}$.\footnote{See \cite{Calabrese:2011vdk,Sotiriadis:2014uza,PhysRevLett.115.157201,Vidmar2016,Pozsgay2017} for some recent work on GGEs and \cite{Langen207,Kinoshita06} for recent experimental realisations. This ensemble was studied from a holographic perspective in \cite{deBoer:2016bov}; see also \cite{Perez:2016vqo}. } 
The trace here is over the Hilbert space of the CFT on a circle.
The dependence of this partition function on the chemical potentials encodes the consequences of Virasoro symmetry for finite-temperature physics in a useful way. The purpose of this paper is to explore the  relation between this ensemble and an  individual microstate of the theory, in the limit of large energies.

The KdV charges are obtained by taking the spatial integral of positive powers of the stress tensor (hence odd-spin) around the spatial circle. The first three are the zero modes of the operators $J_2\equiv T$, $J_4 \equiv (TT)$ and $J_6 \equiv (T(TT)) + \frac{(c+2)}{12} (T'T')$, where $T$ is the stress tensor, the round brackets denote conformal normal ordering, and the prime is a spatial derivative.\footnote{Here we adopt the usual notation where the spatial circle has period $2\pi$, but later it will be convenient to adopt a convention where the spatial circle has period 1.} The additional term in $J_6$ is required to ensure that the zero modes commute, and there will be similar terms in all the higher spin operators. The zero modes of these operators can be written explicitly in terms of the Virasoro modes of the stress tensor. For the first three charges, 
\bea  \label{kdvcharges}
I_1 &\equiv& L_0 - k , \\
I_3 &\equiv& 2 \sum_{n=1}^\infty L_{-n} L_n + L_0^2 - (2k + \frac{1}{6}) L_0 +  k \left(k +\frac{11}{60}\right), \nonumber\\
 I_5 &\equiv& \sum_{n_1+n_2+n_3=0}: L_{n_1} L_{n_2} L_{n_3} :
+\sum_{n=1}^\infty \left( (4k+\frac{11}{6}) n^2 -1 - 6k\right) L_{-n} L_n \nonumber \\  \nonumber
&&
+\frac{3}{2} \sum_{m=1}^\infty L_{1-2m} L_{2m-1} -(3k+\frac{1}{2}) L_0^2 
+\frac{(18k+5)(12k+1)}{72} L_0 - \frac{k(42k+17)(36k+7)}{1512} \,. 
\eea 
Though all the $I_{2m-1}$'s are mutually commuting, and act within each Virasoro module and its level subspace, it is not easy to figure out the basis of descendants where the $I_{2m-1}$'s are diagonal.\footnote{Curiously, such a diagonalization problem can also be mapped to questions in some Schroedinger equations \cite{Bazhanov:2003ni}.}

One motivation for understanding the GGE its importance in the application of the eigenstate thermalization hypothesis (ETH)  \cite{Deutsch1991,PhysRevE.50.888,Rigol2008,Srednicki99,DAlessio2016} to two-dimensional conformal field theories (CFTs), and more broadly in the understanding of chaos in two-dimensional CFTs.\footnote{See \cite{Fitzpatrick:2015zha,deBoer:2016bov,Perez:2016vqo,Lashkari:2016vgj,Dymarsky:2016ntg,Lashkari:2017hwq,Faulkner:2017hll,He:2017txy,Basu:2017kzo,Brehm:2018ipf,Romero-Bermudez:2018dim,Hikida:2018khg} and the references therein for some recent discussions.} ETH is the conjecture that for a suitable class of simple operators, the matrix elements of the operators between generic energy eigenstates at high energy will have a diagonal contribution which matches the expectation values in the microcanonical ensemble at that energy, and exponentially suppressed off-diagonal terms. We can also substitute the canonical ensemble at an appropriate temperature for the microcanonical ensemble; there are then power-law corrections in the temperature from the relation between the canonical and microcanonical ensembles. 

ETH is expected to be valid for theories exhibiting chaotic dynamics. It is known to be violated in integrable theories; intuitively, the special nature of an integrable Hamiltonian implies that the energy eigenstates do not sample state space equally. CFTs are an interesting intermediate case, as they have an infinite-dimensional symmetry algebra, the Virasoro algebra, but unlike in integrable models, the existence of this infinite-dimensional symmetry does not necessarily trivialise the dynamics.\footnote{With the exception of minimal models.} CFTs can exhibit chaotic dynamics \cite{Roberts:2014ifa} and hence could be expected to obey the ETH. 

In \cite{Basu:2017kzo} ETH in two-dimensional CFT was considered, and it was argued that there were operators whose expectation values matched between a typical microstate and the canonical ensemble only at leading order in central charge as $c\to\infty$. However, in the presence of conserved charges ETH is modified; we expect to need to introduce chemical potentials in the thermal ensemble to match the expectation values of the charges in the microstate (intuitively, conserved charges are not expected to thermalise, so their expectation values in energy eigenstates are not universal).  
Indeed, the operators considered in \cite{Basu:2017kzo} are related to the KdV charges.
Once we match the conserved charges the other simple operators are expected to exhibit ETH, in that the matrix elements in energy eigenstates should be related to the expectation value in an ensemble which matches the value of the conserved charge in the eigenstate. In the CFT context, we would expect the expectation values in energy eigenstates to be related to a generalised Gibbs ensemble (GGE) for some particular values of the chemical potentials.

Thus, the question considered in \cite{Basu:2017kzo} is more properly considered as one of ensemble equivalence. The conclusion of that work implies that at subleading order in the central charge, the values of the KdV charges in a microstate cannot be reproduced by the thermal ensemble, that is by the GGE with vanishing chemical potentials. The question is then whether there is a non-zero value of the chemical potentials for which the GGE would successfully reproduce the microstate values. This matching is not in itself a test of ETH; it is a more primitive question about whether there is an ensemble which matches the microstate.\footnote{This is like choosing the temperature to match the energy of the microstate.} If so, ETH would then be a prediction about the correspondence in the values of other observables. 

In this paper, we wish to consider the GGE at high temperature and study the expectation values of the KdV charges in this ensemble, with a view to identifying the appropriate values of the chemical potentials to match to a generic high energy microstate. We consider this first in the limit of large temperature and large central charge, where we can do a saddle-point analysis. The saddle-point analysis is carried out in section \ref{saddle}. The leading order matching of  \cite{Basu:2017kzo}  might lead one to expect that in the leading large $c$ analysis, the appropriate value of the chemical potentials would be zero. But this is not the case; instead we find that the leading large $c$ analysis does not constrain the values of (appropriately rescaled) chemical potentials. 

We then proceed to study the structure of the finite central charge corrections. Away from large $c$, we need to work perturbatively in the chemical potentials. The coefficients in the perturbative expansion of the log of the GGE partition function are the connected correlation functions of the KdV charges in the thermal ensemble. We therefore study the values of these connected correlation functions at high temperature.  

In a companion paper \cite{paper1}, we studied the behaviour of the correlation functions at finite temperature and finite central charge. We found that the correlation functions are differential operators acting on the partition function, and showed that they transform as quasi-modular forms under conformal transformations, drawing on the analysis of  \cite{Dijkgraaf:1996iy}. Here we use this modular transformation to relate the high-temperature behaviour to low temperatures. In section \ref{thermal} we explain the modular transformations and use them together with results from our companion paper to give explicit expressions in a few cases.  

We will see explicitly that the coefficients at subleading order in $k$ have more structure, so the matching will no longer be satisfied for arbitrary chemical potentials. Unfortunately, as noted in \cite{Lashkari:2017hwq}, there is an infinite system of equations to solve for the chemical potentials which requires knowledge of all the correlation functions, so we are not able to determine the required values of the chemical potentials explicitly. 

We also use our results to study the statistics of the eigenvalues of the KdV charges at high energies, and at high level in a given Verma module in section \ref{verma}. The correlation functions of KdV charges restricted to a given Verma module can be evaluated by considering the same differential operator acting on the character. This expression does not have simple modular transformation properties, but we can still evaluate the high temperature behaviour for generic Verma modules, by expressing the character in terms of the Dedekind eta function and making use of the transformation properties of the eta function and the Eisenstein series. We will find that the statistics are sharply peaked; the correlation functions of the charges approximately factorise, with the connected correlation functions being suppressed by powers of the temperature relative to the total correlation function. 

\section{Saddle-point analysis at large central charge}
\label{saddle}

In this section we will consider the GGE in the limit of large temperature and large central charge, where we can compute the partition function directly by a saddle-point analysis.  In later sections we will consider the analysis at finite central charge, which is considerably more intricate.

%

At high temperature, the GGE is dominated by heavy states, with dimension $h \sim c/\beta^2$ and $n \sim 1/\beta^2$.  At finite $c$ the computation of a KdV charge in such a state is quite complicated, but at large central charge, $h \gg n$, and we can approximate $I_{2m-1} \approx L_0^m \approx h^m$. Thus, at leading order at large $c$, the KdV charges take approximately the same value on the descendents contributing to the GGE partition function  
\begin{equation}
Z_{GGE}  = \mbox{Tr}[ e^{-\beta I_1 + \mu_3 I_3 + \mu_5 I_5 \ldots }], 
\end{equation}
and we can approximate this simply by  an integral over the conformal dimension $h$ of the primaries,
\begin{equation}\label{integrand}
Z_{GGE}  = \int dh  e^{2\pi \sqrt{4kh} -\beta h + \mu_3 h^2 + \mu_5 h^3 \ldots }, 
\end{equation}
where we have used Cardy's formula for the density of states at large energies. From this we immediately see that the partition function satisfies 
\begin{equation}
\partial_{\mu_{2m-1}} Z_{GGE}  = (-1)^{m} \partial_\beta^m Z_{GGE}.
\end{equation}

The partition function can be computed in this limit in a saddle point approximation. %
%
%
To simplify our equations, we will define a rescaled dimension $\bar h$ and chemical potentials $\tilde \mu_{2m-1}$ by 
\begin{equation}
h = \frac{(2\pi)^2 k}{\beta^2} \bar h,~~~~~~
\tilde \mu_{2m-1} = \mu_{2m-1} \frac{(2\pi)^{2m-2} k^{m-1}}{\beta^{2m-1} }.
\end{equation}
In the standard thermal ensemble (with all $\mu_{2m-1}$ set to zero) the partition function is dominated by a saddle with $\bar h=1$, so $\bar h$ measures the deviation from the saddle point used to derive Cardy's formula.
%
%
The saddle-point value  of $\bar h$ is found by extremizing the log of the integrand \eqref{integrand}:
\begin{equation}
f({\bar h})  \equiv  \frac{(2\pi)^2 k}{\beta} [2 \sqrt{\bar h} - \bar h  + \tilde \mu_3  \bar h^2 + \tilde \mu_5 \bar h^3 + \ldots]. 
\end{equation}
We note that the saddle point value of $\bar h$ is a function only of the $\tilde \mu_{2m-1}$.

%
The solution of the saddle point equation is non-trivial, so let us begin by just considering the simple case where we keep just $\mu_3$ non-zero.
In this case the saddle point equation
\begin{equation}
\frac{1}{\sqrt{\bar h}} - 1 + 2 \tilde \mu_3 \bar h = 0
\end{equation}
is a cubic equation for $\sqrt{\bar h}$ which can be solved explicitly in terms of radicals.  The simplest way to write the solution is as a hypergeometric function
\begin{equation}
\bar h = \frac{1}{3 \tilde \mu_3} \left[ 1-{}_2 F_1(\frac{1}{3}, - \frac{1}{3}; \frac{1}{2} ;  \frac{27 \tilde \mu_3}{2} ) \right] = 1 + 4 \tilde \mu_3 + 28 \tilde \mu_3^2+  240 \tilde \mu_3^3 + 2288 \tilde \mu_3^4 +  \ldots
\end{equation}
In the final expression we have expanded the result near $\tilde \mu_3=0$, and see that at small $\mu_3$ the saddle reduces to the usual thermal saddle with ${\bar h}=1$ as expected. 
Substituting this back into $f(\bar h)$ to compute the GGE partition function gives 
\begin{eqnarray} \label{GGE3}
\log Z_{GGE}(\mu_3)  &=&  \frac{(2\pi)^2 k}{\beta} \frac{1}{6 \tilde \mu_3} \left[ {}_2 F_1(-\frac{2}{3}, - \frac{1}{3}; \frac{1}{2} ;  \frac{27 \tilde \mu_3}{2} ) -1 \right].
\nonumber \\ \nonumber
&=& \frac{(2\pi)^2 k}{\beta} + \frac{(2\pi)^4 k^2}{\beta^4} \mu_3 + \frac{4(2\pi)^6 k^3}{\beta^7} \mu_3^2 +\frac{24 (2\pi)^8 k^4}{\beta^{10}} \mu_3^3 + \frac{176 (2\pi)^{10} k^5}{\beta^{13}} \mu_3^4 + \mathcal{O}(\mu_3^5).
\end{eqnarray}
Again we have expanded near $\tilde \mu_3=0$, and see that at small $\mu_3$ the saddle point reduces to the usual Cardy formula for the free energy at high temperature.

%
%
When we take all of the chemical potentials to be non-zero we can no longer explicitly solve for the saddle point in terms of radicals.  However, we can still write down an expression for the saddle point values of $\bar h$ as an infinite series expansion in the $\tilde \mu_{2m-1}$,
\begin{equation} 
\bar h = 1 + 2 \sum_m m\,  \tilde \mu_{2m-1} + \sum_{m,n} mn (2m+2n-1) \tilde \mu_{2m-1} \tilde \mu_{2n-1}+  \ldots.  
\end{equation}
Substituting this in gives 
\begin{equation} \label{GGEZ}
\log Z_{GGE} =  
 \frac{(2 \pi)^2 k}{\beta}  \left[ 1 +  \sum_m  \tilde \mu_{2m-1} + \sum_{mn} mn\,  \tilde \mu_{2m-1} \tilde \mu_{2n-1} + \ldots \right].   
\end{equation}
Continuing to arbitrary orders, we have
\begin{equation} \label{GGEZ}
\log Z_{GGE} = 
 \frac{(2 \pi)^2 k}{\beta}  \left[ 1 + \sum_{n=1}^\infty \sum_{m_1 \ldots m_n=2}^\infty a_{m_1 \ldots m_n} \tilde \mu_{2m_1-1} \ldots \tilde \mu_{2m_n-1} \right],   
\end{equation}
where the purely numerical coefficients are 
\begin{equation} \label{as}
a_{m_1 \ldots m_n} = 2 
\frac{(2 M-n)!}{(2 M+2-2n)!n!} \prod_{i=1}^n m_i,
\end{equation}
with $M = \sum_{i=1}^n m_i$. 
%
%
This can be viewed as a generalization of Cardy's formula to include the KdV potentials.

The values of the KdV charges in the GGE are determined by taking derivatives of $\log Z$ with respect to the chemical potentials:
\begin{equation} \label{GGEvev1}
\langle I_{1} \rangle_{GGE} = - \partial_{\beta} \log Z_{GEE}[\beta, \{ \mu_{2m-1} \}], \end{equation}
\begin{equation} \label{GGEvev2}
\langle I_{2m-1} \rangle_{GGE} = \partial_{\mu_{2m-1}} \log Z_{GGE}[\beta, \{ \mu_{2m-1} \}]. 
\end{equation}
The coefficients in the perturbative expansion \eqref{GGEZ} are the connected correlation functions of the KdV charges in the ordinary thermal ensemble (with all of the $\mu_{2n-1}$ set to zero). 
For example, the connected correlation functions of $I_3$ at finite temperature can be read off from the Taylor expansion \eqref{GGE3} around $\mu_3=0$:
\begin{equation}
\langle I_3 \rangle_\beta =  \partial_{\mu_3} \log Z_{GGE}|_{\mu_3=0} = \frac{(2\pi)^4 k^2}{\beta^4}, \quad  \langle I_3^2 \rangle_{\beta, c} =  \partial^2_{\mu_3} \log Z_{GGE}|_{\mu_3=0} = \frac{8(2\pi)^6 k^3}{\beta^7} , \end{equation}
\begin{equation}
\langle I_3^3 \rangle_{\beta,c} =  \partial^3_{\mu_3} \log Z_{GGE}|_{\mu_3=0} = \frac{144 (2\pi)^8 k^4}{\beta^{10}} , \quad \langle I_3^4 \rangle_{\beta,c} =  \partial^4_{\mu_3} \log Z_{GGE}|_{\mu_3=0} = \frac{4224 (2\pi)^{10} k^5}{\beta^{13}}.
\end{equation}
We will return to these connected correlation functions later, when we consider the finite central charge analysis. 

 We can now consider the approximation of a particular microstate by a GGE ensemble. The values of the GGE potentials $\beta$ and $\mu_{2m-1}$ will be determined by demanding that the expectation values of KdV charges match the values in our microstate $|\psi\rangle$:
\begin{equation}
\langle I_{2m-1} \rangle_{\psi} = \langle I_{2m-1} \rangle_{GGE}.
\end{equation}
At large central charge, the typical high energy microstate  will have $h \gg n \gg 1$ as noted previously, so we can approximate the value of $I_{2m-1}$ on the microstate by the $L_0^m$ contribution, 
\begin{equation} \label{micro}
\langle I_{2m-1} \rangle_{\psi} \approx h^m. 
\end{equation}
Matching the microstate values \eqref{micro}  therefore requires $\langle I_{2m-1} \rangle_{GGE} = \langle I_1 \rangle_{GGE}^m$, i.e.
\begin{equation} \label{taufn}
 \partial_{\mu_{2m-1}} \log Z_{GGE}[\beta, \{ \mu_{2m-1} \}]  =  (- \partial_{\beta} \log Z_{GGE} [\beta, \{ \mu_{2m-1} \}] )^m~.
\end{equation}
%
The temperature is fixed by the condition that
\begin{equation} \label{heq}
h   =  - \partial_{\beta} \log Z_{GGE}[\beta, \{ \mu_{2m-1} \}].
\end{equation}
%

We might expect that the first conditions \eqref{taufn} would fix the chemical potentials $\tilde \mu_{2m-1}$ to zero at leading order in large $k$. 
Indeed, if we fix the chemical potentials to zero this gives $-\partial_{\beta} \log Z_{GGE} = \frac{(2\pi)^2 k}{\beta^2}$, and the fact that $a_{m_1} = 1$ implies $\partial_{\mu_{2m-1}} \log Z_{GGE}[\beta, \{ \mu_{2m-1} \}] = \left( \frac{(2\pi)^2k}{\beta^2} \right)^m$. 
However, we now see that at large $k$ we are not required to fix the chemical potentials to zero.  In the large $k$ saddle-point calculation, the condition \eqref{taufn} is {\it automatically} satisfied.  The partition function in the saddle-point approximation is given by $\log Z = f(h_\star)$, where $h_\star$ is the saddle-point value with $f'(h_\star) = 0$. Thus  
\begin{equation} 
- \partial_{\beta} \log Z_{GGE}[\beta, \{ \mu_{2m-1} \}] = h_\star - \frac{\partial f}{\partial h_\star} \frac{\partial h_\star}{\partial \beta} = h_\star, 
\end{equation}
and
\begin{equation} 
 \partial_{\mu_{2m-1}} \log Z_{GGE}[\beta, \{ \mu_{2m-1} \}] = h_\star^m + \frac{\partial f}{\partial h_\star} \frac{\partial h_\star}{\partial \mu_{2m-1}} = h_\star^m, 
\end{equation}
and \eqref{taufn} is satisfied independent of the choice of chemical potentials. 

Our conclusion is that the leading large $k$ analysis does not fix the values of the chemical potentials.  This does not mean the values of the chemical potentials are irrelevant to the leading order analysis; the relation between the temperature and the energy of the microstate in \eqref{heq} depends on the values of the chemical potentials. 


At subleading orders in $k$ the values of the KdV charges for a typical microstate are not simply determined by the energy; we need to consider contributions from the non-zero Virasoro generators in the KdV charges. Thus the values of the chemical potentials will not be uinversal. Since the temperature depends on the choice of the values of the chemical potentials through \eqref{heq}, it will also not be universal; it is proportional to $\sqrt{h}$, but the multiplicative factor depends on the particular values of the chemical potentials.  

In the next sections, we will explore the subleading in $k$ corrections to this saddle-point analysis, by using modular transformations to calculate the connected correlation functions of the KdV charges at high temperatures. We will see explicitly that for example with just $\mu_3$ non-zero, $\langle I_3 \rangle_{GGE} \neq \langle I_1 \rangle_{GGE}^2$ once we include subleading corrections.  There will be non-trivial dependence on the chemical potentials which could in principle be used to match the values of the KdV charges in a particular microstate. 

Note that as is clear from the leading order expression \eqref{GGEZ}, the partition function depends on all the chemical potentials equally, and the value of $\langle I_3 \rangle_{GGE}$ will depend on the values of all the $\tilde \mu_{2m_i-1}$, not just $\tilde \mu_3$. Thus, as noted in \cite{Lashkari:2017hwq}, solving for the chemical potentials to match the values of the KdV charges in a particular microstate would require us to solve an infinite system of equations.  We will nevertheless be able to determine many features of the finite $k$ ensemble quite explicitly. 

\section{Correlation functions at  finite $c$ by modular transformation}
\label{thermal}

At finite central charge we cannot determine the exact GGE partition function using a saddle-point analysis.  We will instead adopt a different approach. We will calculate the partition function perturbatively in the chemical potentials $\mu_i$ by evaluating the thermal correlation functions of the KdV charges.  This will give us the coefficients in the Taylor expansion of $\log Z_{GGE}$ in the $\mu$'s.  Our general strategy will be to study the theory on a circle, where one can use modular transformations to determine the correlation functions at high temperature in terms of low temperature correlators which are easy to compute. 

The modular transformation properties of the correlators can be determined using an argument of \cite{Dijkgraaf:1996iy}, which we will now review and extend.  The basic idea is as follows.  The correlator of a set of conserved charges at finite temperature is given by the integrals of correlation functions of the corresponding currents on the torus:
\begin{equation}
\langle I_{2m_1-1} \ldots I_{2m_n-1} \rangle = \langle \oint J_{2m_1} \ldots \oint J_{2m_n} \rangle 
\end{equation}
These objects do not transform covariantly under modular transformations because the choice of integration around the "spatial" circle in the contour integrals fixes a basis of cycles on the torus. However, since the currents are conserved the position of the contour integrals does not matter, so we can relate these contour integrals to surface integrals over the entire torus.  The only subtlety is that there will be contributions from singularities where the operators coincide, but these ``anomaly" terms can also be expressed as surface integrals of operators over the torus.  Since a surface integral over the torus does not involve a choice of spatial circle, the surface of a current will transform covariantly under modular transformations.  Thus using the expression in terms of surface integrals we can easily map from high to low temperatures by a modular transformation, and then re-express the surface integrals in terms of contour integrals which can be explicitly evaluated at low-temperature.  

We now describe this procedure more explicitly.
Our torus has a spatial cycle of period $1$ and a time cycle of period $\frac{\beta}{2\pi}$, so the modular parameter is $\tau = i \frac{\beta}{2\pi}$ and $q = e^{2\pi i \tau} = e^{-\beta}$. We relate high temperatures ($\beta \to 0$) to low temperatures by the modular transformation $\hat \tau = - \tau^{-1}$. 
We will adopt the notation of \cite{Dijkgraaf:1996iy}, where $\oint J = \int_0^1 \frac{dz}{2\pi} J(z)$ and $\int J = \int \frac{d^2z}{2 \pi \tau_2} J(z)$. 

The one-point functions of KdV charges are then 
\begin{equation}
\langle I_{2m-1} \rangle (q) = \langle \oint J_{2m} \rangle (q) = \langle \int J_{2m} \rangle (q) = \frac{1}{\tau^{2m}}   \langle \int J_{2m} \rangle (\hat q) = \frac{1}{\tau^{2m}}   \langle \oint J_{2m} \rangle (\hat q) =  \frac{1}{\tau^{2m}} \langle I_{2m-1} \rangle (\hat q),
\end{equation}
where in the first two steps we expressed the charge as the integral of the current and then used the fact that the current is conserved to smear the integral over the torus.  A factor of $\tau_2$ appears in the definition of $\int J_{2m}$ because it is the area of the torus. In the next step we use the fact that the one point function transforms like a modular form of weight $2m$ to relate the high-temperature behaviour to low temperature, which we then rewrite as a contour integral to express the result in terms of the low-temperature one-point function of the KdV charge. The result is that the one point function transforms like a modular form of weight $2m$.  

The low-temperature one-point function is dominated by the contribution of the vacuum state, up to corrections that are exponentially small in $\tilde q$. The expectation values of the KdV charges in the vacuum state are trivially calculated from the definitions; for example $ \langle 0| I_3 | 0 \rangle = k (k + \frac{11}{60} )$, $ \langle 0| I_5 | 0 \rangle = - k \left( k + \frac{17}{42} \right) \left( k + \frac{7}{36} \right)$.  This leads to the high temperature expectation values
\begin{equation} \label{I3high}
\frac{\langle I_{3} \rangle}{Z}  (q) \approx \frac{1}{\tau^4} k \left( k + \frac{11}{60} \right) +\dots = 
 \frac{(2\pi)^4}{\beta^4}  k \left( k + \frac{11}{60} \right) + \dots,
\end{equation}
\begin{equation}
\frac{\langle I_{5} \rangle}{Z}  (q) \approx - \frac{1}{\tau^6} k \left( k + \frac{17}{42} \right) \left( k + \frac{7}{36} \right) +\dots = 
\frac{(2\pi)^6}{\beta^6}  k \left( k + \frac{17}{42} \right) \left( k + \frac{7}{36} \right) + \dots 
\end{equation}
where the corrections $\dots$ are exponentially suppressed (as $e^{-{\cal O}(\beta^{-1})}$) as $\beta\to0$.

For higher point functions life gets more interesting.  The reason is that when we convert a product of contour integrals into a product of surface integrals we must worry about the points where the operators in the surface integrals coincide.  These singular terms can be analyzed by looking at the OPE of the operators appearing in the surface integrals.  For example, following \cite{Dijkgraaf:1996iy} we can evaluated the two point function of a pair of charges on the torus to obtain 
\begin{equation}
\langle \int J_{2m} \int J_{2n} \rangle = \langle \oint J_{2m} \oint J_{2n} \rangle +  \frac{1}{2\tau_2} \langle \oint [J_{2m} J_{2n}]_2 \rangle.
\end{equation}
Here the $[J_{2m} J_{2n}]_2$ is the operator which appears as the singular term of order $(z-w)^{-2}$ in the OPE $J_{2m}(z) J_{2n}(w)$, and we use a square bracket notation to avoid confusion with the round bracket notation for conformal normal ordering. 
We will refer to this term a  "modular anomaly term," since it will imply that the correlation function does not transform as a modular form but rather as a quasi-modular form.

We can apply this to write the two-point functions of KdV charges in terms of quantities which transform covariantly
\begin{eqnarray}
\langle I_{2m-1} I_{2n-1} \rangle &=& \langle \oint J_{2m} \oint J_{2n} \rangle  = \langle \int J_{2m} \int J_{2n} \rangle - \frac{1}{2\tau_2} \langle \int [J_{2m} J_{2n}]_2 \rangle \\ \nonumber
& =& \frac{1}{\tau^{2(m+n)}}  \langle \int J_{2m} \int J_{2n} \rangle(\hat q) - \frac{1}{2\tau_2} \frac{1}{\tau^{2(m+n-1)}} \langle \int [J_{2m} J_{2n}]_2 \rangle(\hat q) \\ \nonumber
&=&  \frac{1}{\tau^{2(m+n)}}[ \langle I_{2m-1} I_{2n-1} \rangle(\hat q) + \frac{1}{2 \hat \tau_2} \langle \oint [J_{2m} J_{2n}]_2 \rangle (\hat q) ] - \frac{1}{2\tau_2} \frac{1}{\tau^{2(m+n-1)}} \langle \oint [J_{2m} J_{2n}]_2 \rangle(\hat q),
\end{eqnarray}
where in the second line we applied the modular transformation and then in the third line we rewrote the expression in terms of contour integrals to obtain an expression we can evaluate. Using $\tau = i \tau_2$, $\hat \tau_2 = \tau_2^{-1} = i \tau^{-1}$ we obtain
\begin{equation}
\langle I_{2m-1} I_{2n-1} \rangle (q) = \frac{1}{\tau^{2(m+n)}}[ \langle I_{2m-1} I_{2n-1} \rangle(\hat q) - i \tau  \langle \oint [J_{2m} J_{2n}]_2 \rangle (\hat q) ].
\end{equation} 
At high temperature the right hand side is dominated by the contribution of the vacuum state, up to exponentially small corrections.  

The important point is that the vacuum state is unique, so the correlation function $\langle I_{2m-1} I_{2n-1} \rangle(\hat q) = \langle I_{2m-1}\rangle (\hat q)\langle I_{2n-1} \rangle(\hat q)+\dots $ factorizes.  Our modular transformation argument then implies that at high temperature the correlation functions will approximately factorize into a product of one point functions as well.  In other words, the statistics of the charges will become sharply peaked at high temperature.  

To make this more explicit, let us consider the connected correlation function: 
\begin{eqnarray}
\langle I_{2m-1} I_{2n-1} \rangle_{\beta, c} &\equiv& \frac{\langle I_{2m-1} I_{2n-1} \rangle}{Z} - \frac{\langle I_{2m-1} \rangle \langle I_{2n-1} \rangle}{Z^2} \\ \nonumber
& =& \frac{1}{\tau^{2(m+n)}} \left[ \frac{\langle I_{2m-1} I_{2n-1} \rangle(\hat q)}{Z} - \frac{\langle I_{2m-1} \rangle(\hat q) \langle I_{2n-1} \rangle (\hat q)}{Z^2} - i \tau \frac{ \langle \oint [J_{2m} J_{2n}]_2 \rangle (\hat q)}{Z} \right] 
\end{eqnarray}
The first two contributions will cancel up to exponentially suppressed corrections, and the connected correlation function is 
\begin{equation} \label{twopt}
\langle I_{2m-1} I_{2n-1} \rangle_{\beta, c} = -i  \frac{1}{\tau^{2(m+n)-1}} \frac{ \langle \oint [J_{2m} J_{2n}]_2 \rangle (\hat q)}{Z} +\dots.
\end{equation}
where $\dots$ denotes corrections that are exponentially suppressed.
There are two important observations to make about this result.  The first is that it is suppressed by one power of the temperature relative to the full correlator, which scales as $\tau^{-2(m+n)}$.  This reproduces the temperature dependence seen in the saddle point analysis, and implies that the correlation functions approximately factorize at high temperature.   The second is that all of the corrections to this factorization which are power law (as opposed to exponentially) suppressed at high temperature are determined entirely in terms of the modular anomaly term evaluated in the vacuum state.
These two features will persist for the higher correlation functions as well.

As an example let us evaluate this explicitly for $\langle I_3^2 \rangle_{\beta ,c}$. We have\footnote{The last total derivative term is in fact irrelevant, as it will not affect the value of the charge obtained when we do the contour integral.}
\begin{equation} \label{j4j4}
[J_4 J_4]_2 = [(TT)(TT)]_2 = 8 (T(TT)) + 8 (T'T') + (24k+10) (T'' T) +( 6k+ \frac{1}{6}) T''''. 
\end{equation}
Taking the zero mode and evaluating this on the vacuum state we find   
\begin{equation}
\langle I_3^2 \rangle_{\beta, c}  = \frac{8 (2\pi)^6}{\beta^7}  k \left(k+ \frac{11}{60} \right) \left( k + \frac{37}{84} \right) + \dots
\end{equation}
where the $\dots$ are exponentially suppressed at high temperature.

We can apply the methods of \cite{Dijkgraaf:1996iy} to higher point functions as well.  For three point functions we obtain
\begin{eqnarray}
\langle \int J_{2m} \int J_{2n} \int J_{2p} \rangle
&=&
\langle \oint J_{2m} \oint J_{2n} \oint J_{2p} \rangle
\\ \nonumber
&&+\frac{1}{2\tau_2} 
\langle
 \oint  J_{2m} \oint [J_{2n} J_{2p}]_2 
+ \oint  J_{2n} \oint [J_{2p}J_{2m}]_2 
+ \oint  J_{2p} \oint [J_{2m} J_{2n}]_2 
\rangle
\\ \nonumber
&&+\frac{1}{2(2\tau_2)^2} 
\langle
 \oint  [J_{2m}  [ J_{2n} J_{2p}]_2 ]_2
+ \oint  [J_{2n}  [ J_{2p} J_{2m}]_2]_2 
+ \oint  [J_{2p}  [J_{2m} J_{2n}]_2]_2 
\rangle.
\end{eqnarray}
This gives the transformation 
\begin{eqnarray}
\langle I_{2m-1} I_{2n-1} I_{2p-1} \rangle &=& \frac{1}{\tau^{2(m+n+p)}} \left[ \langle I_{2m-1} I_{2n-1} I_{2p-1} \rangle(\hat q) \right. \\ \nonumber
&&-  i \tau  \langle
 I_{2m-1} \oint [J_{2n} J_{2p}]_2 
+ I_{2n-1} \oint [J_{2p}J_{2m}]_2 
+ I_{2p-1} \oint [J_{2m} J_{2n}]_2 
\rangle(\hat q) \\ \nonumber
&& \left. - \frac{\tau^2}{2} \langle
 \oint  [J_{2m}  [ J_{2n} J_{2p}]_2 ]_2
+ \oint  [J_{2n}  [ J_{2p} J_{2m}]_2]_2 
+ \oint  [J_{2p}  [J_{2m} J_{2n}]_2]_2 
\rangle (\hat q)  \right]
\end{eqnarray}
In the connected three-point function the leading and first subleading terms cancel due to the factorization of the low-temperature correlation functions, teaving us with 
\begin{eqnarray}
\langle I_{2m-1} I_{2n-1} I_{2p-1} \rangle_{\beta,c} &\equiv& \frac{ \langle I_{2m-1} I_{2n-1} I_{2p-1} \rangle}{Z} - \frac{  \langle I_{2m-1}\rangle \langle I_{2n-1} I_{2p-1} \rangle}{Z^2}  -  \frac{\langle I_{2n-1}\rangle \langle I_{2p-1} I_{2m-1} \rangle}{Z^2}  \\ \nonumber 
&& - \frac{ \langle I_{2p-1}\rangle \langle I_{2m-1} I_{2n-1} \rangle}{Z^2} + 2 \frac{  \langle I_{2m-1}\rangle \langle I_{2n-1}\rangle \langle I_{2p-1} \rangle}{Z^3} \\ \nonumber 
&=& -
\frac{1}{2\tau^{2(m+n+p)-2}} \frac{\langle
 \oint  [J_{2m}  [ J_{2n} J_{2p}]_2 ]_2
+ \oint  [J_{2n}  [ J_{2p} J_{2m}]_2]_2 
+ \oint  [J_{2p}  [J_{2m} J_{2n}]_2]_2 
\rangle (\hat q) }{Z}.
\end{eqnarray}
We see that the connected three-point function at high temperature is determined entirely by the most inhomogeneous term in the transformation of $\langle I_{2m-1} I_{2n-1} I_{2p-1} \rangle$.
In addition, as in the case of the two point function, we see that the correlation functions approximately factorize at high temperature, since the three point function is suppressed by two powers of the temperature relative to the total correlation function.
Evaluating this explicitly for $\langle I_3^3 \rangle_{\beta,c}$ we find
\begin{equation}
\langle I_3^3 \rangle_{\beta ,c}  = \frac{144 (2\pi)^8}{\beta^{10}} k \left( k+ \frac{11}{60} \right) \left( k^2 + \frac{41}{30} k + \frac{257}{720} \right) +\dots
\end{equation}
where the $\dots$ are exponentially suppressed.

We can continue to higher orders.  For example, at fourth order we have 
\begin{multline} 
 \langle ( \int J_{2m})^4 \rangle =  \langle (\oint J_{2m})^4 \rangle + \frac{6}{2 \tau_2} \langle (\oint J_{2m})^2 \oint [J_{2m}J_{2m}]_2 \rangle + \frac{6}{8\tau_2^2} \langle \oint [J_{2m}J_{2m}]_2 \oint [J_{2m}J_{2m}]_2 \rangle\\ + \frac{12}{8 \tau_2^2} \langle \oint J_{2m} \oint [J_{2m}[J_{2m}J_{2m}]_2]_2 \rangle  + \frac{8}{32 \tau_2^3} \langle \oint [J_{2m}[J_{2m}[J_{2m}J_{2m}]_2]_2]_2 \rangle \\+ \frac{4}{32 \tau_2^3} \langle \oint [[J_{2m}J_{2m}]_2[J_{2m}J_{2m}]_2]_2 \rangle,
\end{multline}
where to make the formula more compact we have considered just the case of four identical currents.  This gives the transformation rule 
\begin{eqnarray}
\langle I_{2m-1}^4 \rangle &=& \frac{1}{\tau^{8m}} \left[ \langle I_{2m-1}^4 \rangle(\hat q) - 6 i \tau \langle I_{2m-1}^2 \oint [J_{2m} J_{2m}]_2 \rangle(\hat q)  - 3 \tau^2 ( \langle \oint [J_{2m} J_{2m}]_2 \oint [J_{2m} J_{2m}]_2 \rangle (\hat q) \right. \nonumber \\ &&+  
 2  \langle  I_{2m-1} \oint [J_{2m}[J_{2m} J_{2m}]_2]_2 \rangle (\hat q) )   + i \tau^3   ( \langle \oint [[J_{2m} J_{2m}]_2  [J_{2m} J_{2m}]_2]_2 \rangle (\hat q)  \\ &&+  \nonumber \left. 
  2  \langle  \oint [J_{2m}[J_{2m}[J_{2m} J_{2m}]_2]_2]_2 \rangle (\hat q) ) \right] 
\end{eqnarray}
We obtain a similar cancellation at high temperatures in the connected correlation function:\footnote{We note that this connected correlation function is a cumulant of the distribution of KdV charges; at fourth order there is a difference between the cumulant and the central moment of a distribution.  The lower-order contributions do not fully cancel if we consider just the central moment.}
\begin{eqnarray}
\langle I_{2m-1}^4 \rangle_{\beta ,c} &\equiv& \frac{\langle I_{2m-1}^4 \rangle}{Z} - 4 \frac{\langle I_{2m-1} \rangle \langle I_{2m-1}^3 \rangle}{Z^2} - 3 
\frac{\langle I_{2m-1}^2 \rangle^2}{Z^2} + 12 \frac{\langle I_{2m-1}^2 \rangle \langle I_{2m-1} \rangle^2}{Z^3}  -6 \frac{ \langle I_{2m-1} \rangle^4}{Z^4} \nonumber \\  
&=& \frac{i}{\tau^{8m-3} Z}  ( \langle \oint [[J_{2m} J_{2m}]_2  [J_{2m} J_{2m}]_2]_2 \rangle (\hat q)  + 2  \langle  \oint [J_{2m}[J_{2m}[J_{2m} J_{2m}]_2]_2]_2 \rangle (\hat q) ). 
\end{eqnarray}
For example, 
\begin{equation}
\langle I_3^4 \rangle_{\beta ,c}  = \frac{4224 (2\pi)^{10}}{\beta^{13}} k \left( k+ \frac{11}{60} \right) \left( k^3 + \frac{817}{220} k^2 + \frac{6439}{2640} k + \frac{43291}{95040} \right) + \dots
\end{equation}
Again, only the most inhomogeneous term contributes to the connected correlator, and is suppressed by three powers of the temperature relative to the total correlation function. 

The cancellations in the connected correlation functions will continue to all orders, so that 
\begin{equation} \label{connfinite}
\langle I_{2m_1-1} \ldots I_{2m_n-1} \rangle_{\beta,c} \approx \frac{1}{\beta^{2M -n +1}} C_{m_1 \ldots m_n}(k) +\dots   
\end{equation}
where $M = \sum_{i=1}^n m_i$, and the coefficient $C_{m_1 \ldots m_n}$ is a function of the central charge.  The $\dots$ corrections to this expression are exponentially suppressed at high temperature. 

It is useful to compare these results to the large central charge saddle point analysis of the previous section.  The first thing to note is that the 
${\beta^{-2M +n -1}}$ in equation (\ref{connfinite}) precisely reproduces the temperature dependence obtained at large central charge.  Indeed, it is also possible to compare the coefficient as well.
This is because at large central charge the leading contribution to $C_{m_1 \ldots m_n}(k)$ comes from a product of stress tensors with no derivatives. Thus to determine the leading large $k$ behaviour we only need to determine the coefficient of the product of stress tensors in the modular anomaly term. Calculating these coefficients at leading order is relatively straightforward, even for arbitrary combinations of KdV charges.  This analysis precisely reproduces the results of the saddle point analysis in the previous section, so we will not go into details here.

It is also useful to compare this analysis with than in our companion paper \cite{paper1}, where we show that the finite-temperature correlation functions of KdV charges can be written as differential operators acting on the partition function. In that analysis, the inhomogeneous part of the transformation of the correlation functions comes from explicit factors of the Eisenstein series $E_2$ appearing in the differential operator.  Each factor of $E_2$ is multiplied by a differential operator, which is precisely the thermal expectation value of one of the modular anomaly terms.  For example, in the differential operator for $\langle I_{2n-1} I_{2m-1}\rangle$ the coefficient of $E_2$ is a differential operator which, when applied to a partition function, computes the thermal one point function $\oint [J_{2m} J_{2n}]_2 \rangle$.  

We can now return to our expression (\ref{connfinite}) for the connected correlation function. 
These connected correlation functions can be used to compute the GGE partition function at finite central charge.  The explicit perturbative expressions derived give us this partition function perturbatively in the chemical potentials.  For example, with just $\mu_3$ turned on we can use our explicit results for $I_3$ to obtain\footnote{As with the large central charge analysis, it is not an accident that $\mu_3$ always appears with $\beta$ in combinations of $\mu_3/\beta^3$. This is guaranteed by dimensional analysis in the thermodynamic limit.  The point is that if we were to consider a circle of size $R$ rather than a circle of size $1$, then we would have to take $\beta \to \beta/R$ and $\mu_3\to \mu_3/R^{3}$, since $I_3$ has scaling dimension $3$.  The only combination which survives the thermodynamic limit is $\mu_3/\beta^3$.

It is also not an accident that each term in this expansion has a factor of $k(k+{11\over 60})$.  The KdV charge $I_3$ vanishes in both the $(2,3)$ and $(2,5)$ minimal models, i.e. in the trivial $c=0$ theory and in the Lee-Yang theory with $k=-{11\over 60}$.  Thus the $\mu_3$-dependent terms in the GGE partition function must vanish as well at these values of the central charge. Similar factors will appear for other terms in the GGE partition function.}
\begin{eqnarray} 
\log Z_{GGE}(\mu_3) &=& \frac{(2\pi)^2 k}{\beta} + \frac{(2\pi)^4}{\beta^4}  k \left(k+ \frac{11}{60} \right) \mu_3 + \frac{4(2\pi)^6}{\beta^7} k \left(k+ \frac{11}{60} \right) \left( k + \frac{37}{84} \right) \mu_3^2 \nonumber \\&&  + \frac{24 (2\pi)^8}{\beta^{10}} k \left( k+ \frac{11}{60} \right) \left( k^2 + \frac{41}{30} k + \frac{257}{720} \right)\mu_3^3  \\ &&+\frac{176(2\pi)^{10}}{\beta^{13}}k \left(k+\frac{11}{60} \right) \left(k^3+\frac{817}{220} k^2+\frac{6439}{2640}k+\frac{43921}{95040} \right)\mu_3^4+  \mathcal{O}(\mu_3^5). \nonumber
\end{eqnarray}
This gives the corrections to the large central charge expression \eqref{GGE3}.
We can see explicitly that 
\begin{equation}
\langle I_3 \rangle_{GGE}  = - \partial_{\mu_3}\log Z_{GGE} \neq (- \partial_\beta \log Z_{GGE})^2
\end{equation}
demonstrating that the universality of the large $k$ results  does not persist at finite central charge.

\section{Statistics at high temperature in a single Verma module} 
\label{verma}

In the last section we studied the statistics of KdV charges using modular transformations.  This allowed us to compute the high temperature behaviour of the connected correlation functions of the KdV charges at high temperature.  We learned that the connected correlators are suppressed relative to the disconnected piece by powers of the temperature, implying that the distribution of KdV charges are sharply peaked; the variance (and all higher cumulants) of the distribution are parametrically suppressed relative to the mean.  

In this section we will derive analogous results for the statistics of the KdV charges within a given Virasoro representation.  We will see that even within a Virasoro representation, the statistics of KdV charges on high level descendant states are sharply peaked.  We can no longer use modular invariance to make this argument, as modular invariance mixes up different Virasoro representations.  We will therefore use the results of our companion paper \cite{paper1}, where the correlation functions of KdV charges within a representation was shown to be a (quasi-modular) differential operator acting on the character of the representation. 

We will consider $c>1$ CFTs, where the Virasoro representations are (except for the vacuum representation) Verma modules.  The Verma module character is
\begin{equation}\label{chiV}
\chi_{h}=\frac{q^{h-k}}{\prod_{n=1}^{\infty}(1-q^n)},
\end{equation}
We will denote the thermal expectation value of a KdV charge within this representation as
\be
\langle I_{2m_1-1}\dots I_{2m_n-1} \rangle_h \equiv {\rm Tr}_h \left[ q^{L_0} I_{2m_1-1}\dots I_{2m_n-1}\right]
\ee
These can be computed as differential operators applied to the character \eqref{chiV}. 
For example, 
\begin{equation} \label{expp}
\langle I_3 \rangle_h = \left[ \partial^2 -{1\over 6} E_2 \partial +{k \over 60} E_4\right] \chi_h,
\end{equation}
where $\partial = q \partial_q$ and $E_{2n}$ is an Eisenstein series.  We refer to Appendix C of \cite{paper1} for a more complete list of differential operators.

It is easy to evaluate the derivative of the character \eqref{chiV}: 
\begin{equation}
\partial \frac{1}{\prod_{n=1}^{\infty}(1-q^n)} =  \frac{1}{\prod_{n=1}^{\infty}(1-q^n)} \sum_{n=1}^\infty \frac{ n q^n}{1-q^n} = \frac{1}{\prod_{n=1}^{\infty}(1-q^n)}  \frac{1}{24} (1-E_2), 
\end{equation}
so that
\begin{equation} \label{expp}
\partial \chi_{h}= \left[ h-k + \frac{1}{24} - \frac{E_2}{24} \right] \chi_h = \left[\tilde h -  \frac{E_2}{24} \right] \chi_h,
\end{equation}
Here for future convenience we have introduced a shifted level $\tilde h \equiv h-k+\frac{1}{24}$ to simplify our formulae. 
The result is
\begin{equation} \label{I3exp}
\langle I_3 \rangle_h = \left[ \tilde h^2 - \frac{1}{4} \tilde h E_2 + \frac{E_2^2}{192} + \left( \frac{k}{60} + \frac{1}{288} \right) E_4 \right] \chi_h. 
\end{equation}

Under modular transformations, we have $E_4(-1/\tau) = \tau^4 E_4(\tau)$, $E_6(-1/\tau) = \tau^6 E_6(\tau)$, $E_2(-1/\tau) = \tau^2 E_2(\tau) + \frac{6 \tau}{\pi i}$, which give the high temperature behaviour 
\begin{equation} \label{eiss}
E_2 \approx - \left( \frac{2\pi}{\beta} \right)^2 + \frac{12}{\beta}, \quad E_4 \approx  \left( \frac{2\pi}{\beta} \right)^4, \quad E_6 \approx -  \left( \frac{2\pi}{\beta} \right)^6, 
\end{equation}
up to exponentially suppressed corrections.  
Using \eqref{eiss} in \eqref{I3exp},  we find
\begin{equation} \label{I3h}
\frac{\langle I_3 \rangle_h}{\chi} = \tilde h^2 + \frac{\pi^2}{\beta^2} \tilde h - \frac{3 \tilde h}{\beta} +  \frac{(25+48k) \pi^4}{180 \beta^4} - \frac{\pi^2}{2\beta^3} + \frac{3}{4 \beta^2} + \dots
\end{equation}
where the $\dots$ denote terms which are exponentially suppressed at high temperature.
We note that if we take $\beta\to 0$ while holding $\tilde h$ fixed this goes like $\beta^4$, just like the result in a full CFT \eqref{I3high}.  But the coefficient is different, and is suppressed by one order of the central charge.  

We can nevertheless reproduce the correct result for the full partition function by summing this over all representations in the theory:
\begin{equation}
\langle I_3 \rangle=  \int dh \left(\frac{1}{\sqrt{2(h-k+\frac{1}{24})}} e^{2\pi \sqrt{\frac{(24k-1)}{6}(h-k+\frac{1}{24})} } \right)\langle I_3 \rangle_h,
\end{equation}
where the expression in parenthesis is the Cardy formula for the density of states of primary operators of dimension $h$ in a CFT with $c>1$.  This density of states can be derived by looking at the modular transformation properties of the partition function which counts primary states (note that we have been careful to keep the power-law correction to the usual exponential factor). At high temperature the character is
%
\begin{equation} \label{chihighT}
\chi(q) \approx e^{\frac{\pi^2}{6 \beta} - \beta \tilde h} \sqrt{ \frac{\beta}{2\pi}}, 
\end{equation}
up to exponentially suppressed corrections. Evaluating the integral at the saddle point (which is at $h_\star=(k-1/24)\frac{(2\pi)^2}{\beta^2}$) gives
\begin{equation}
\langle I_3 \rangle/Z
\approx \frac{(2\pi)^4}{\beta^4}  k \left( k + \frac{11}{60} \right) + \dots. 
\end{equation} 
exactly reproducing our previous result.  We note that it is necessary to carefully keep track of the subleading terms in the saddle point analysis in order to see that all of the potential power law corrections cancel and that the $\dots$ in this formula are indeed exponentially suppressed in $\beta$. 
It is important to note that the order $k^2$ part of this result, which dominates at large central charge, comes just from the ${\cal O}(h^2)$ contribution in $\langle I_3\rangle_h$. This is consistent with the expectation that the contribution from the conformal dimension of the primary determines the high temperature behaviour at large central charge. 

For the higher correlation functions, we find cancellations in the connected correlation functions, just as in the previous section. Using the differential operator for $\langle I_3^2\rangle_h$ from \cite{paper1} we have
\begin{equation}
\begin{aligned} \label{I3sqexp}
\langle I_3^2 \rangle_h &= \left[ \tilde h^4 - \frac{1}{2} \tilde h^3 E_2  - \frac{5}{96} \tilde h^2 E_2^2 + \frac{ 24k+95}{720} \tilde h^2 E_4 + \frac{25}{1152} \tilde h E_2^3 - \frac{(7+24k)}{360} \tilde h E_6+ \frac{168k-19}{2880} \tilde h E_2 E_4 \right.  \\
&\left. -\frac{5}{12288}E_2^4-\frac{19+120k}{46080}E_2^2E_4+\frac{(19536 k (12 k+5)+12425)}{14515200}E_4^2+\frac{(7-24 k (120 k+29))}{181440}E_2 E_6
\right] \chi_h.
\end{aligned}
\end{equation}
Evaluating this expression in the high temperature limit using \eqref{eiss}, there will be a contribution which goes like $\beta^{-8}$ from the terms in the second line. When we consider the connected correlation function, there are cancellations. If we consider first finite temperature, the expression is most cleanly given in terms of derivatives of the Eisenstein series, 
\begin{multline}
\frac{\langle I_3^2 \rangle_h}{\chi} - \frac{\langle I_3 \rangle^2_h}{\chi^2}  = - \frac{3}{2} \partial E_2 \tilde h^2 + \left( \frac{7}{4} \partial^2 E_2 + \frac{48k+49}{240} \partial E_4 \right) \tilde h\\  - \frac{1}{8} \partial^3 E_2 - \frac{(211+192k)}{9600} \partial^2 E_4 - \frac{(1883+6816k+11520 k^2)}{362880} \partial E_6.
\end{multline}
Evaluating this expression in the high temperature limit, the most divergent term comes from the last term, $\partial E_6 \sim \beta^{-7}$, so 
\begin{equation}
\frac{\langle I_3^2 \rangle_h}{\chi} - \frac{\langle I_3 \rangle^2_h}{\chi^2} = \frac{(1883 + 6816 k + 11520 k^2) \pi^6}{945 \beta^7}  +
\dots, 
\end{equation}
where the subleading terms include subleading powers of $\beta$, which we can determine as in \eqref{I3h}, but have not written explicitly for simplicity. We see that the connected correlation function is suppressed by one power of $\beta$ relative to the full correlation function,  just as in the previous section when we computed correlation functions in the full CFT. 

Similarly, $\langle I_3^3 \rangle_h$ involves contributions going like $\beta^{-12}$, but in the connected correlation functions the first two orders cancel, giving 
\begin{equation}
\frac{\langle I_3^3 \rangle_h}{\chi} - 3 \frac{\langle I_3 \rangle_h \langle I_3^2 \rangle_h}{\chi^2} + 2 \frac{\langle I_3 \rangle_h^3}{\chi^3} = \frac{(25925 + 135504 k + 407808 k^2 + 552960 k^3) \pi^8}{225 \beta^{10}}  + \dots 
\end{equation}
where again the subleading terms include subleading powers of $\beta$. 

Our conclusion is that the distribution of eigenvalues of the KdV charges is sharply peaked, with the variance and higher cumulants being suppressed by powers of the temperature relative to the mean.  
This is a novel result: at low temperatures the correlation functions factorize and the primary state dominates because the KdV charges are simply powers of $L_0$. But at high temperatures the character is dominated by descendants with large level (much larger than $h$).  In this limit the $L_0^m$ term scales like $1/\beta^{2m}$ with a numerical ($k$-independent) coefficient.  The $k$-dependent part of the expectation value comes from the other terms in the KdV charge, making the cancellations in the connected correlators (and hence the fact that the distribution of KdV charges is sharply peaked) non-trivial. 

Finally, we note that we are calculating here a thermal average over all the states in the Verma module, but at high temperature the calculation is dominated by a narrow range of levels with $n \approx \frac{\pi^2}{6 \beta^2}$.  This is the usual equivalence between canonical and microcanonical ensembles at high temperature.  In fact, the statistics of KdV charges at a particular level can be computed exactly (see section 7 of \cite{paper1}); the results agree with the canonical ensemble computation described here.

\section*{Acknowledgements}
We thank Anatoly Dymarsky, Jan de Boer, Nima Lashkari and Jan Manschot for discussions. G. N. is supported by Simons Foundation Grant to HMI under the program ``Targeted Grants to Institutes''. SFR is supported by STFC under consolidated grants ST/L000407/1 and ST/P000371/1. I.T. is supported by the Alexander S. Onassis Foundation under the contract F ZM 086-1.
A.M. acknowledges the support of the Natural Sciences and Engineering Research Council of Canada (NSERC), funding reference number SAPIN/00032-2015.
This work was supported in part by a grant from the Simons Foundation (385602, A.M.).

\bibliographystyle{utphys}
\bibliography{kdvDraft1}

\end{document}